\documentclass[journal]{IEEEtran}
 
\usepackage[hyphenbreaks]{breakurl}

\ifCLASSINFOpdf
  \usepackage[pdftex]{graphicx}
  
\else
  
\fi

\usepackage{caption}
\usepackage{subcaption}
\usepackage{algorithmic}

\PassOptionsToPackage{hyphens}{url}
 \usepackage[dvipsnames]{xcolor}
 \usepackage{soul}
 \usepackage[normalem]{ulem}
 \usepackage{makecell}

\usepackage[utf8]{inputenc}
\usepackage{cite}
\usepackage{float}
\usepackage[autostyle]{csquotes}
 
 \usepackage[cmex10]{amsmath}
\usepackage{esvect}
\usepackage{amssymb}
\usepackage{graphicx}
 
 \usepackage[hidelinks]{hyperref}
 \usepackage[noabbrev,capitalize]{cleveref}

\usepackage{tikz}
\usepackage{pgfplotstable}
\usepackage{pgfplots}
\usepackage{threeparttable}
\usepackage{multicol,mathtools, lipsum}
\usepackage{ulem}
\usepackage{relsize}
\usepackage{todonotes}

\usepackage[linesnumbered,ruled,vlined]{algorithm2e}

\SetKwInput{KwInput}{Input}                
\SetKwInput{KwOutput}{Output}              

\SetCommentSty{mycommfont}
\usepackage{url}
\usepackage{multirow}
\usepackage{lscape}
\usepackage{cuted}
 \usepackage{mwe}
\usepackage{upgreek}
\DeclareMathOperator*{\argmin}{arg min}
\usepackage{placeins}
\usepackage{lipsum,amsmath,multicol}

\usepackage{stfloats}

\usepackage{adjustbox}
\usepackage{enumitem}
\usepackage{siunitx}
\usepackage{booktabs}
\usepackage{diagbox}
 
\pgfplotsset{compat=newest} 

\makeatother
\makeatother

\title{A Federated Deep Learning Approach for Privacy-Preserving Real-Time Transient Stability Predictions in Power Systems}
 \author{
 \IEEEauthorblockN{Maeshal~Hijazi and  Payman~Dehghanian}

 \IEEEauthorblockA{Department of Electrical and Computer Engineering\\
 The George Washington University\\
 Washington, DC 20052, USA.\\
$\{$mahijazi, payman$\}$@gwu.edu}

\thanks{This work was supported by the Government of the Kingdom of Saudi Arabia and in part by the George Washington University (GWU)'s University Facilitating Fund (UFF) Grant.}
}

\begin{document}
\maketitle

\begin{abstract}
Maintaining the privacy of power system data is essential for protecting sensitive information and ensuring the operation security of critical infrastructure. Therefore, the adoption of centralized deep learning (DL) transient stability assessment (TSA) frameworks can introduce risks to electric utilities. This is because these frameworks make utility data susceptible to cyber threats and communication issues when transmitting data to a central server for training a single TSA model. Additionally, the centralized approach demands significant computational resources, which may not always be readily available. In light of these challenges, this paper introduces a federated DL-based TSA framework designed to identify the operating states of the power system. Instead of local utilities transmitting their data to a central server for centralized model training, they independently train their own TSA models using their respective datasets. Subsequently, the parameters of each local TSA model are sent to a central server for model aggregation, and the resulting model is shared back with the local clients. This approach not only preserves the integrity of local utility data, making it resilient against cyber threats but also reduces the computational demands for local TSA model training. The proposed approach is tested on four local clients each having the IEEE 39-bus test system.  

\end{abstract}
\vspace{-4mm}
\section{Introduction} \label{Intro}
\noindent 
The ongoing transformation and digitization of the power grid, facilitated by various sensing and actuating technologies such as phasor measurement units (PMUs), smart inverter-based resources (IBRs), etc. have significantly increased the volume of data available across the power grid \cite{wang2021distributed}. While this wealth of data is valuable for enhancing the accuracy of deep learning (DL) models used in transient stability assessment (TSA), training these centralized models demands substantial computational resources that not all utility providers may possess. Furthermore, these emerging technologies introduce vulnerabilities to cyber-attacks and data communication delays and failures, which could result in inaccurate decisions and at times severe power outages \cite{hijazi2022spatio}. They also introduce uncertainties in electricity flow, heightened control complexity, and potential threats to system stability. In response to these emerging challenges, there is a growing need for an efficient, computationally feasible, and well-protected real-time TSA platform capable of effectively assessing the operational state of the power system.

In recent years, there has been a notable shift towards the utilization of centralized DL models consisting of networks such as convolutional neural networks (CNNs) and long short-term memory (LSTM). These networks are often combined with other DL techniques like active learning (AL), transfer learning (TL), and Bayesian modeling to predict the transient stability of the power system. For instance, reference \cite{wang2022power} aims to improve TSA in power systems by integrating various advanced techniques, including DL, Bayesian modeling, and AL. Reference \cite{shi2020convolutional} employs a CNN to create a classifier that determines whether a power system is in a stable, aperiodic unstable, or oscillatory unstable condition. It does so by analyzing brief snapshots of the voltage measurements collected from PMUs right after a fault occurs. In reference \cite{james2017intelligent}, an LSTM network is utilized to construct a dynamic temporal model for TSA, where the primary goal is to strike a balance between response time and accuracy in assessing the stability of the system. In \cite{azhar2022development}, a method based on a combination of CNN and LSTM networks, i.e., ConvLSTM, is introduced, which was designed to identify transient instability in a power system by analyzing time series data collected from PMUs. Reference \cite{senyuk2023power} introduces a data-driven approach to dynamic stability analysis that can adapt to changes in power system parameters and structure. It uses ensemble machine learning (ML) algorithms to develop emergency control strategies, improving the accuracy of TSA with a focus on low-inertia power systems. In \cite{hijazi2023transfer}, a model-based TL approach is presented that integrates ConvLSTM to address the challenge of variations in topological configurations in power grids and the impacts on the stability assessment accuracy. However, all these references utilize centralized DL models which (1) are computationally intensive, (2) are vulnerable to cyber adversaries, and (3) do not feature privacy-preserving characteristics. 

Federated learning (FL) is a DL technique that is utilized to tackle the above-mentioned challenges in centralized DL models. FL is primarily centered around the idea that it facilitates the training of ML models across distributed data sources \cite{mcmahan2017communication}. This approach prioritizes the preservation of data privacy and security while also being computationally efficient, making it well-suited for applications where data cannot be easily centralized or transmitted to a central server similar to the case for TSA in power systems where data could be owned by different entities (e.g., electric utilities). The literature on utilizing FL for TSA applications in the power grid is scarce. Reference \cite{ren2023secfedsa} proposes a secure federated stability assessment (SecFedSA) method which is a distributed approach based on a differential privacy (DP) mechanism. In addition, a quantum-secured FL system for smart grid dynamic security assessment (QFDSA) is proposed in \cite{ren2023qfdsa} which integrates a quantum key distribution, FL, and dynamic key generation to create a more secure, privacy-preserving, and adaptable system for data-driven dynamic security assessment (DSA) in smart cyber-physical power grids. However, both of these approaches tackle the DSA in the power system, which focuses on the long-term stability of the system, unlike the TSA which focuses on the short-term stability performance, i.e., the assessment of a power system's ability to maintain stability during and immediately following a significant disturbance. In this paper, we propose an FL approach that trains an efficient, computationally feasible, privacy-preserving, and decentralized real-time TSA model capable of effectively assessing the operating state of the power system.

The rest of the paper is structured as follows: Section \ref{Sec:Background} introduces a general background on centralized and federated DL-based TSA models. Section \ref{Sec:ProposedFramework} introduces the proposed framework consisting of the federated DL-based TSA approach and the system stability characterization scheme. Section \ref{sec:results} presents the data acquisition, data-preprocessing, and the architecture of the neural network followed by the numerical assessment of the proposed method. Finally, conclusions are provided in Section \ref{CON}.

\begin{figure*}[!ht]
    \centering
    \includegraphics[width = 2\columnwidth]{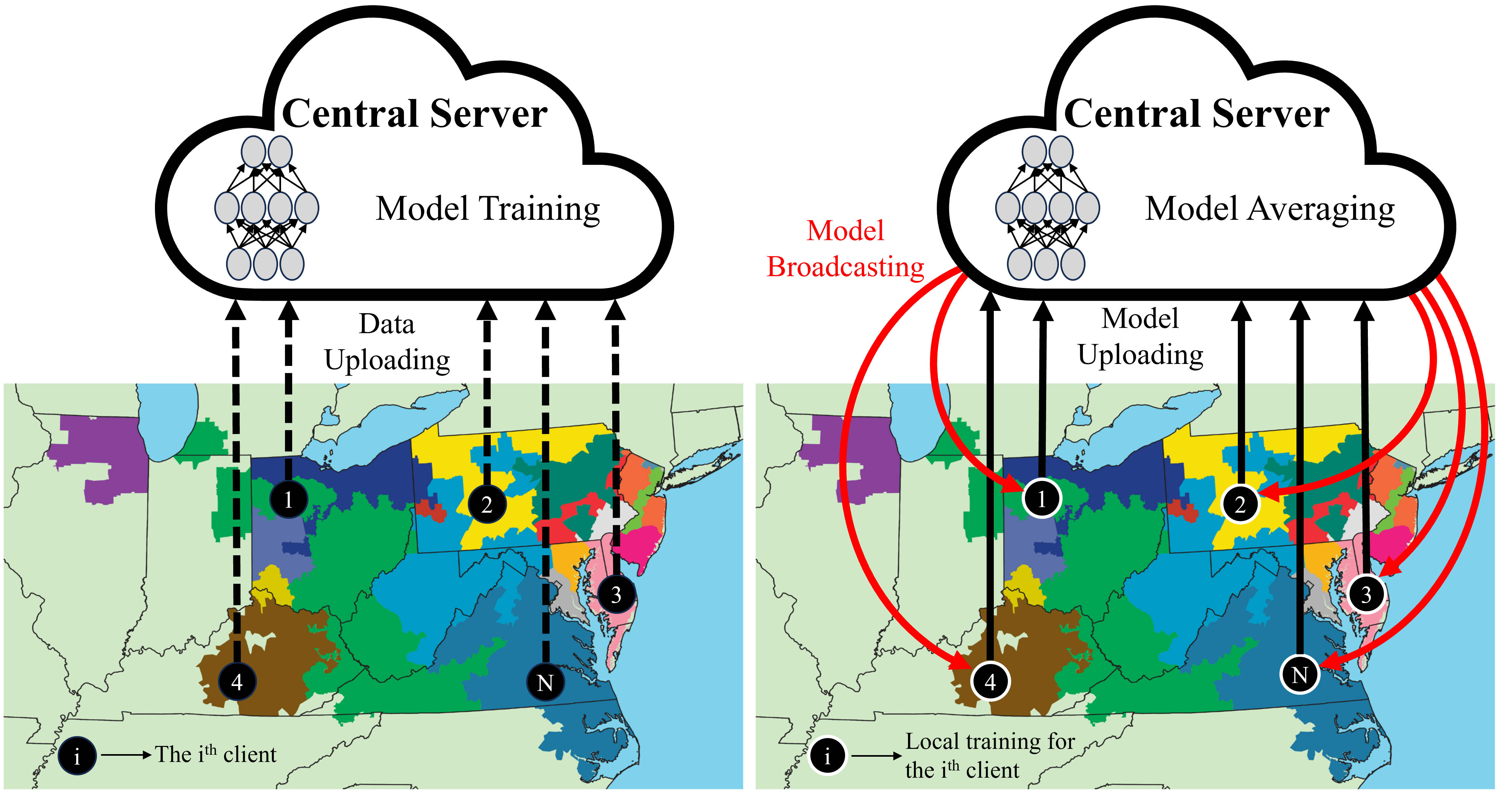}
    \caption{Centralized and federated DL-based TSA model schemes: The big picture.} \label{fig:CenvsFed}
\end{figure*}



\section{Background} \label{Sec:Background}
\vspace{-2mm}
\subsection{Centralized DL-based TSA Models}
In a centralized DL-based TSA model, TSA data is collected from various electric utilities across the power grid and sent to a central server. The database is composed of input features, $\textbf{x}$, which are physical features related to the power system such as bus voltage and current, system frequency, or the active and reactive powers of the generators, and the stability output, $y$, discussed later in Section \ref{sec:DataAcquisition}. The central server then tries to map a relationship between $\textbf{x}$ and $y$, using what is called model parameters or $\boldsymbol{\uptheta}$. The learned DL-based TSA model, known as $g_{\boldsymbol{\uptheta}}(\cdot)$, is formulated as:
\begin{equation}\label{eq:g_theta_dot}
  g_{\boldsymbol{\uptheta}}(\cdot)=g^{(i)}(\cdots g^{(2)}(g^{(1)}(\textbf{x})))
\end{equation}
where $g^{(i)}$ represents the function of the $i$-th layer of a neural network. The whole idea behind  the training process of $g_{\boldsymbol{\uptheta}}(\cdot)$ is to minimize the
difference between the ground truth label $y$ and the predicted $g_{\boldsymbol{\uptheta}}(\textbf{x})$ as:
\begin{equation}\label{eq:loss_function_cen}
  \min_{\boldsymbol{\uptheta}} L_g(g_{\boldsymbol{\uptheta}}(\textbf{x}), y)
\end{equation}
where $L(\cdot)$ represents the loss function of $g_{\boldsymbol{\uptheta}}(\cdot)$. One way to solve this minimization problem in neural networks is 
through gradient descent algorithms to update the DL-based TSA model parameters as:
\begin{equation}\label{eq:update}
\boldsymbol{\uptheta}_{j+1}=\boldsymbol{\uptheta}_{j}-\eta\cdot\nabla_{\boldsymbol{\uptheta}}L_g(g_{\boldsymbol{\uptheta}}(\textbf{x}), y)
\end{equation}
where $\eta$ denotes the learning rate, $\boldsymbol{\uptheta}_{j}$ represents the learned parameters, and $j$ represents the $j$-th iteration step.

\subsection{Federated DL-based TSA Models}
In contrast to centralized DL-based TSA models, federated DL-based TSA models operate in a decentralized manner. That is, data from different electric utilities or agents doesn't have to be entirely shared with a central server. Instead, each agent independently trains their own model using their data and available computational resources. Subsequently, each agent only transmits the parameters of its DL-based TSA models to the central server. This approach ensures the preservation of data privacy and enhances the security of the training process, as only the model weights of each agent are shared with the central server.

An FL system consists of $N$ clients and a centralized server with the dataset $\mathcal{D}_n$ at the $n$-th client, $n=1, 2,\cdots,N$. For all clients, the purpose is to learn a global TSA model over data that resides at the $N$ clients. Thereby, the optimization objective of an FL model can be formulated as:
\begin{equation}\label{eq:loss_function_fed}
  \boldsymbol{\uptheta^*} = \argmin_{\boldsymbol{\uptheta}}g(\mathcal{D};\boldsymbol{\uptheta})
\end{equation}
where $g(\mathcal{D};\boldsymbol{\uptheta})$ is the global objective function and $\mathcal{D}=\{\mathcal{D}_n\}_{n=1}^{N}$. Figure \ref{fig:CenvsFed} shows the different approaches that the centralized and federated DL-based TSA frameworks take.
\vspace{-2mm}
\section{The Proposed Framework}\label{Sec:ProposedFramework}
The proposed framework relies on a combination of FL and DL. In this setup, each client utilizes its individual transient stability database to train a local model, leveraging the computational resources available locally. In this paper, the proposed framework assumes that all clients share a common system, and employ identical neural network architectures for their local models. Subsequently, each client transmits its acquired TSA model parameters to a central server, which then conducts a global TSA 
model averaging. This approach ensures the privacy of data and adherence to data protection regulations, all while enabling the realization of a decentralized TSA. Even though a client could rely on training a DL model based on a wide variety of system settings in an offline manner, not all clients would have the computational resources to be able to train a DL model covering all power system configurations and dynamics. The following subsections will discuss the detailed procedures for the proposed federated DL-based TSA model framework and the detailed system stability characterization schemes for real-time TSA.

\begin{figure*}[t]
    \centering\includegraphics[width=2\columnwidth]{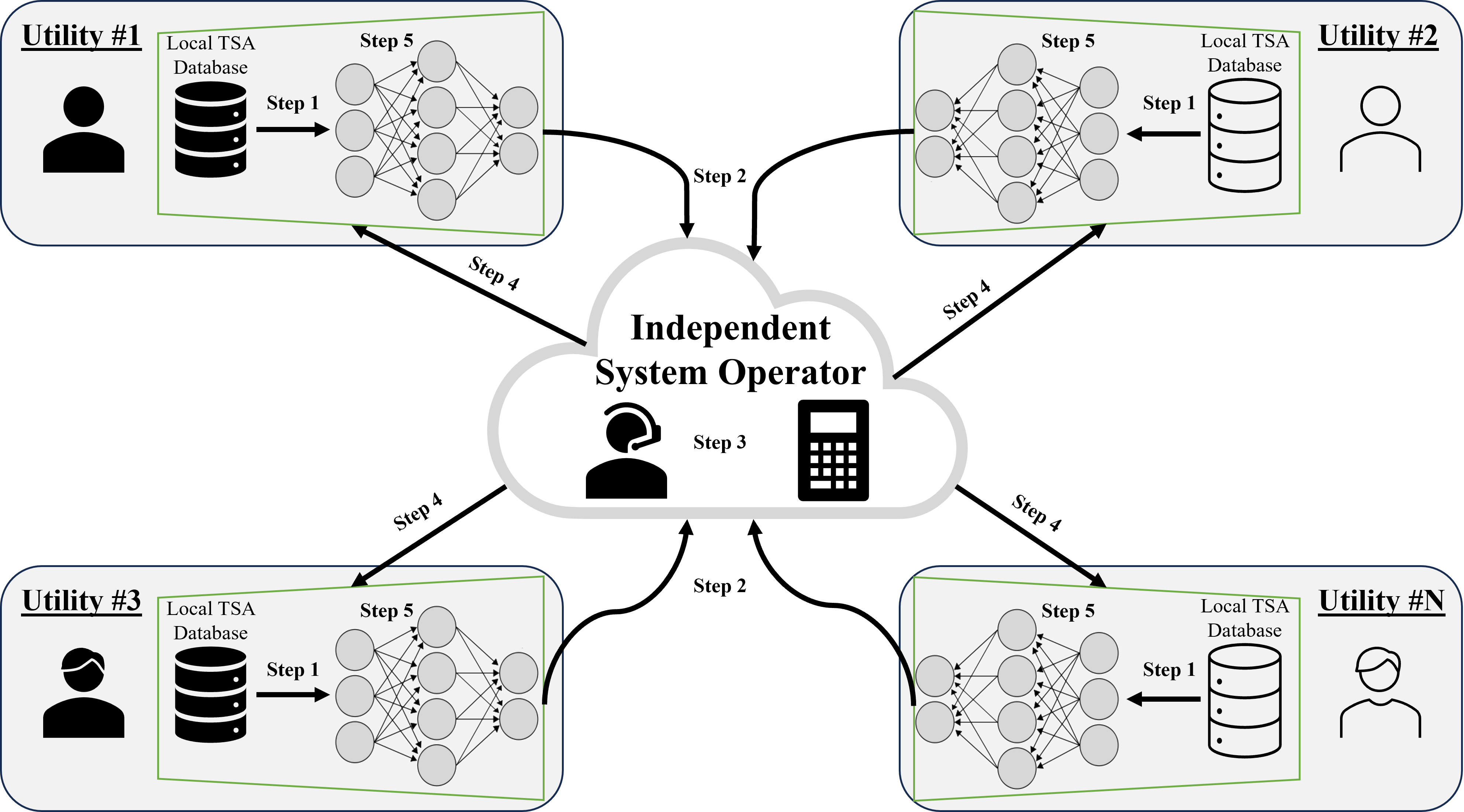} 
    \caption{Proposed federated DL-based TSA model framework.} \label{fig:framwork}
\end{figure*}
\vspace{-4mm}
\subsection{Procedures for the Federated DL-based TSA Framework}
The detailed procedure for the proposed federated DL-based TSA framework is described as follows:
\begin{itemize}
    \item \textbf{Step 0: \textit{Initialization}.} In this step, the global TSA model parameters are initialized and distributed to each client, along with the number of communication rounds, $\mathcal{C}$, between local clients and the central server.

    \item \textbf{Step 1: \textit{Update of the Local TSA Models}.} Once each of the $N$ clients receives the initialized or broadcasted global TSA model parameters from the central server, they fine-tune those parameters using their local database of input system features and their corresponding output, $\mathcal{D}_n$. The goal for each client is to find a minimum loss function, $g_{\boldsymbol{\uptheta}}(\cdot)$, based on their local sample database. 

    \item \textbf{Step 2: \textit{Upload of the Local TSA Models onto the Client Server}.}
    Local TSA models trained by each of the $N$ clients are uploaded to the central server. For example, learned model parameters for client $i$ and for the $c$-th communication round are given as $\boldsymbol{\uptheta^c_i}$.

    \item \textbf{Step 3: \textit{Global TSA Model Aggregation}.}
    In this step and once all local clients have sent their local TSA model parameters for a given communication round to the central server, the central server performs model averaging of all local TSA model parameters or what is known as federated averaging \cite{konevcny2016federated}. Federated averaging for the global TSA model parameters for the $c$-th communication round is calculated as:
    \begin{equation}\label{eq:fed_avg}
  \boldsymbol{\uptheta_G^c}=\frac{\sum_{i=1}^{N}\boldsymbol{\uptheta_i^c}}{N}
    \end{equation}
where, $N$ is the number of local clients.

    \item \textbf{Step 4: \textit{Global TSA Model Broadcast}.}
    In this step, the global TSA model parameters for the $c$-th communication round, $\boldsymbol{\uptheta_G^c}$, are broadcasted back to all local clients. 


    \item \textbf{Step 5: \textit{Test of the Local TSA Models}.}
    Once the global TSA model parameters,  $\boldsymbol{\uptheta_G^c}$, have been broadcasted to the local clients, they are tested using the local client testing data. If the testing criteria are satisfied, the training process is considered complete. However, if the criteria are not met, the process recommences from \textbf{Step 1}.
\end{itemize}

In summary, the process of updating and transmitting models between the central server and all the clients necessitates multiple communication rounds. Each round consists of an uploading phase, during which the clients upload their local TSA models, and a downloading phase, in which the server consolidates the TSA model and disseminates it to the clients. Subsequently, the clients adjust their local TSA models based on the global TSA model. Please refer to Algorithm \ref{alg} and Fig. \ref{fig:framwork} for further clarification.

\begin{algorithm}[!h]
    \caption{The Proposed Federated DL-based TSA Framework}\label{alg}
    \KwInput{$N$ local clients with their corresponding TSA dataset $\mathcal{D}_n=\{(\textbf{x}_{n,d},y_{n,d})\}^{|\mathcal{D}_n|}_{d=1}$ at the $n$-th client, the learning rate $\eta$, total communication rounds $\mathcal{C}$.}
    
    \KwOutput{The trained global TSA model, $\boldsymbol{\uptheta_G^\mathcal{C}}$}

    $\textbf{\textit{Initiliaze}:}$ Set the initial global TSA model parameters to $\boldsymbol{\uptheta_G^0}$, at $c=0$ \tcp*{\textbf{Step 0}}

    The central server broadcasts $\boldsymbol{\uptheta_G^0}$ and $\mathcal{C}$ to all $N$ clients \tcp*{\textbf{Step 0}}
    
    \For{$c=1$ \KwTo $\mathcal{C}$}{
        \tcp{Local TSA Model Training}
        \For{$n=1$ \KwTo $N$}{
            Update local TSA model gradients as $\nabla_{\boldsymbol{\uptheta_{G}^c}}L_{gn}(g_{n}(\textbf{x}_{n,d}), y_{n,d})$

            \tcp{\textbf{Step 1}}
            Update local TSA weight parameters as $\boldsymbol{\uptheta_n^{c+1}}=\boldsymbol{\uptheta_n^{c}}-\eta\cdot\sum_{d=1}^{|\mathcal{D}_n|}\nabla_{\boldsymbol{\uptheta_{G}^c}}L_{gn}(g_{n}(\textbf{x}_{n,d}), y_{n,d})$
        }
         \tcp{\textbf{Step 2}}
        Upload of local TSA models to the central server
        
        \tcp{Aggregation of the Global TSA Model}
        Calculate the global TSA model parameters, $\boldsymbol{\uptheta_G^{c+1}}$, using \ref{eq:fed_avg} \tcp*{\textbf{Step 3}}

        The central server broadcasts $\boldsymbol{\uptheta_G^{c+1}}$ to all $N$ local clients \tcp*{\textbf{Step 4}}

        \tcp{Testing Process for all $N$ Local Clients}

        \For{$n=1$ \KwTo $N$}{
        Test the broadcasted global TSA model parameters $\boldsymbol{\uptheta_G^{c+1}}$ with each local TSA database $\mathcal{D}_n$ \tcp*{\textbf{Step 5}}
        }

        \tcp{Update the Iteration of the Communication Rounds}
        $c\leftarrow c+1$
    }
\end{algorithm}
\vspace{-3.5mm}
\subsection{The Proposed System Stability Classification Scheme}\label{sec:TransientStability}
The state of the system following a contingency can theoretically be ascertained using a Transient Stability Index (TSI), which is formally defined as \cite{hijazi2023transfer}:
\begin{equation}\label{eq:eta}
    \eta = \frac{360^\circ-|\Delta\delta|_{max} }{360^\circ+|\Delta\delta|_{max} }
\end{equation}
where  $\Delta\delta_{\small \mbox{max}}$ represents the maximum separation in rotor angles between any two generators that occurs after the fault. The TSI serves as a rapid method for evaluating the system's overall stability performance, relying on Time-Domain Simulations \cite{hijazi2023transfer}, \cite{james2017intelligent}.
The results of the simulations are used to categorize the system's stability profiles as either stable or unstable, depending on the value of $\eta$. A system is characterized as stable only when $\eta$ exceeds 0. The system's operational states are then sorted into five distinct classes, each delineated by a range of underlying events, as elaborated below:
\begin{itemize}
    \item \textbf{Class 1: \textit{Stable Operating State}.} All the observed data matrices either belong to the pre-fault operating time or reveal a stable operating state during the post-fault clearance period.
    \item \textbf{Class 2: \textit{Fault Occurrence}.} Any observed data matrix that covers the instant of fault occurrence.
    \item \textbf{Class 3: \textit{Fault Duration}.} All the observed data matrices cover exactly the timestamps that lie between fault occurrence and fault clearance (without timestamps of either fault occurrence or fault clearance).
    \item\textbf{Class 4: \textit{Fault Clearance}.} Any observed data matrix covers the instant timestamp of fault clearance.
    \item \textbf{Class 5: \textit{Unstable Operating State}.} All observed data matrices containing the instant of unstable operating states and all timestamps afterward during the post-fault clearance period.
\end{itemize} 

The training data is generated as discussed in Section \ref{sec:DataAcquisition}, and is classified and labeled accordingly as detailed above. It is then used to train the federated DL-based TSA model presented in Section \ref{Sec:ProposedFramework}.

\vspace{-4mm}
\section{Numerical Tests and Analyses} \label{sec:results}

\subsection{Training Data Acquisition}\label{sec:DataAcquisition}
The parameters utilized for the training of the proposed federated DL-based TSA model encompass various electrical attributes, including current and voltage magnitudes, rotor angle, voltage phase angles, and system frequency. The training data for each local service territory is sourced from their respective PMUs located at all generator buses within their network. This data serves to disclose comprehensive system information and is derived from TSA simulations conducted on the IEEE 39-bus test network using the PowerWorld simulation environment \cite{PW}. The IEEE 39-bus test system consists of 39 buses, 10 generating units, 31 load points, and 34 transmission lines. In this paper, it is assumed, without loss of generality, that four distinct operators are managing the same IEEE 39-bus test system in their service territories. The TSA simulation settings carried out for all clients are summarized in Table \ref{Tbl:TSASim}. Each simulation has a duration of 20 seconds, utilizing a time-step of 0.0167 seconds, resulting in 1200 time-stamped recordings for each contingency scenario. For each contingency, a fault is introduced precisely at $t$ = 1 second and is simulated for 16 cycles, while the test system is presumed to operate at a frequency of 60 Hz.
\vspace{-2mm}
\begin{table}
\caption{TSA Simulation Settings for Each Client}
\centering
\setlength\arrayrulewidth{1pt}
\begin{tabular}{c c c}
\hline
\textbf{Client \#} & \textbf{Load Percentage} & \textbf{3$\phi$ Fault Location} \\ 
\hline
1 & Base \& +1\% Loads & Bus\\
2 & -3\% \& +5\% Loads & 25\% of line lengths\\
3 & -2\% \& +3\% Loads & 50\% of line lengths\\
3 & -1\% \& +2\% Loads & 75\% of line lengths\\
\hline
\end{tabular}
\label{Tbl:TSASim}
\end{table}



\begin{figure}
    \centering
    \includegraphics[width=6.5cm]{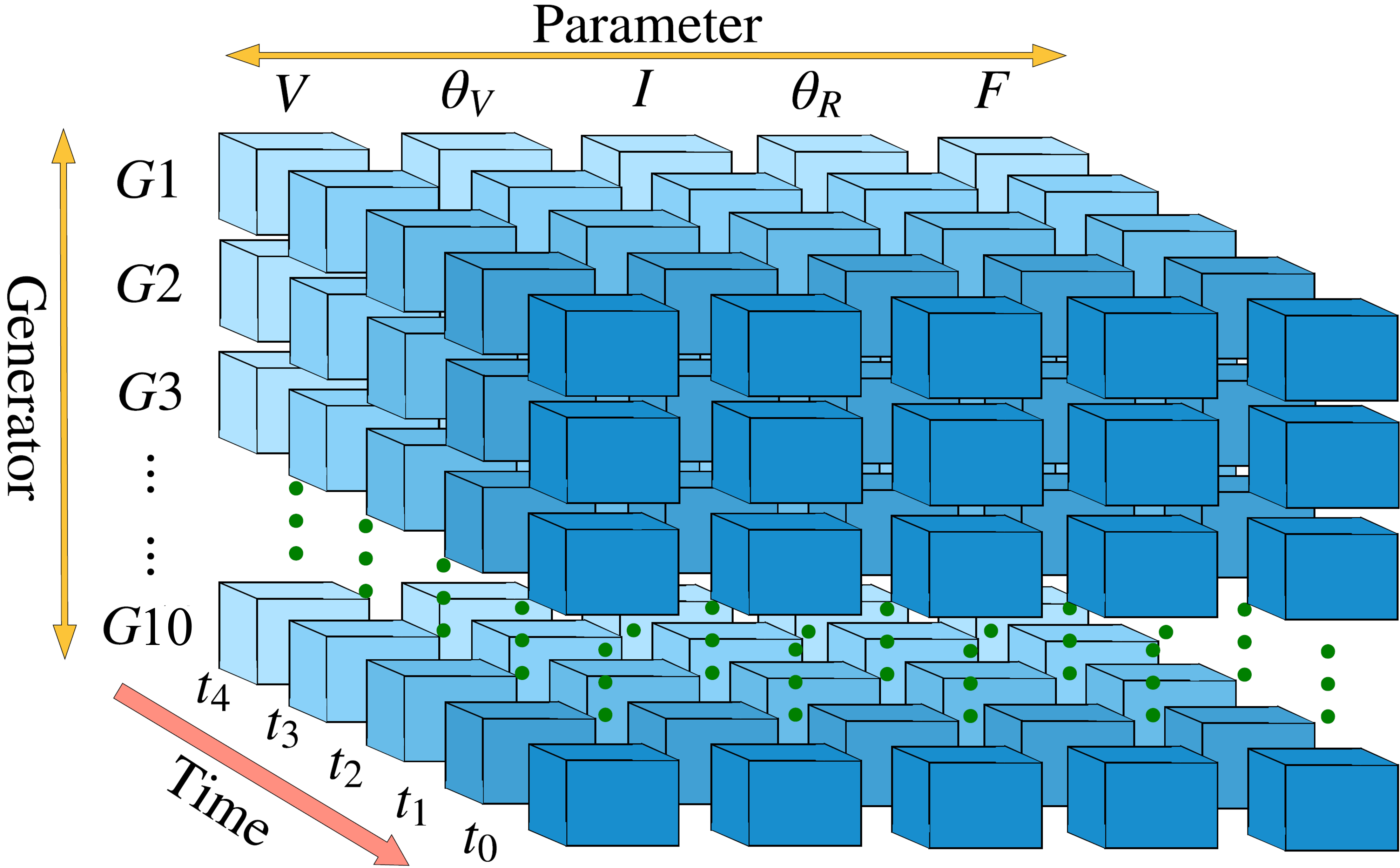}
    \caption{3D data matrix representation for the IEEE 39-bus test system used in the proposed framework.}
    \label{fig:DataStructure}
\end{figure}

\begin{figure*}[t]
    \centering
    \includegraphics[width=2\columnwidth]{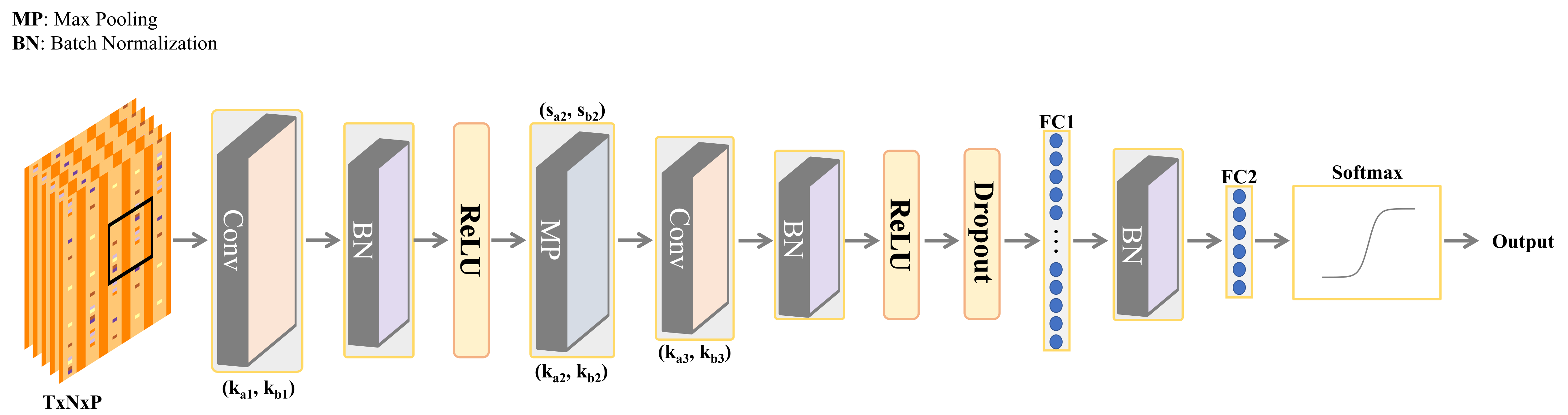}
    \caption{Network architecture for local and global TSA models.}
    \label{fig:network}
\end{figure*}
\vspace{-4mm}
\subsection{Data Pre-Processing}\label{sec:DataPre-processing}
To enable real-time monitoring of the system's transient stability, the surveillance system must continually analyze power grid parameters over consecutive time steps. All parameters detailed in Section \ref{sec:DataAcquisition} are observed within a sliding window of size $t$ time stamps. Consequently, at each sampling instance, this sliding window comprises $(t-1)$ past measurement recordings and one current measurement. The observed raw data is restructured and converted into a three-dimensional vector, representing time steps, generator numbers, and parameters, as depicted in Fig. \ref{fig:DataStructure}. The color shading within this vector is assigned to reflect the passage of time, with the darkest shade indicating the most recent time-step entry in the data matrix. For the IEEE 39-bus test system, there are 10 generators as shown on the \textit{Generator} axis and 5 parameters on the \textit{Parameter} axis. Also, a range of time stamps exists on the \textit{Time} axis. The length of each observation window is 5 time-stamps and the sliding step is 1 time-stamp \cite{hijazi2023transfer}. A heat-map image of the three-dimensional matrix is created for each sample, i.e., the data matrix for each sample is rendered as a color image of size $T \times N \times P$, wherein $T$ is the length of the observation window, $N$ is the number of generators, and $P$ is the number of parameters. Therefore, the size of each heat-map image for any particular fault scenario is considered constant and it is 5$\times$10$\times$5.

\vspace{-4mm}
\subsection{Network Architecture and Settings}\label{sec:architecture}
The neural network architecture for the local and global TSA models is shown in Fig. \ref{fig:network}. The network is composed of 2 main CNN layers, a max pooling layer, and 2 fully connected layers. The kernel sizes used for this network are $(k_{a1}, k_{b1})$ = $(3, 1)$, $(k_{a2}, k_{b2})$ = $(1, 3)$, and $(k_{a3}, k_{b3})$ = $(3, 1)$. Furthermore, the stride kernel for the maximum pooling layer is $(S_{a2}, S_{b2})$ = $(2, 1)$, with a dropout percentage taken to be 20$\%$. 
\vspace{-2mm}
\begin{table}[b]
\caption{TSA Models Parameters}
\centering
\setlength\arrayrulewidth{1pt}
\begin{tabular}{c c}
\hline
\textbf{Parameters} & \textbf{Value} \\ 
\hline
Epochs for local models & 8 \\
Communication rounds $\mathcal{C}$  & 5\\
Learning rate & 0.0003\\
\hline
\end{tabular}
\label{Tbl:ModelParams}
\end{table}
\subsection{Test Results}
\begin{figure*}
     \centering
     \begin{subfigure}[t]{0.25\textwidth}
         \centering
         \includegraphics[width=\textwidth]{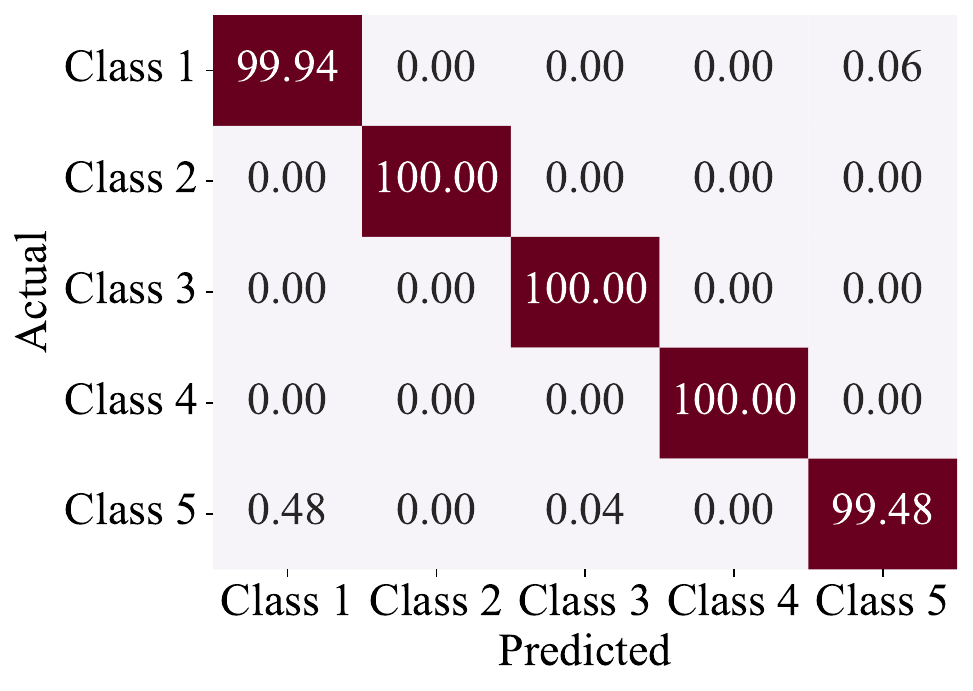}
         \caption{Client \#1}
         \label{fig:Heatmap1}
     \end{subfigure}
     \hspace{-8mm}
     \hfill
     \begin{subfigure}[t]{0.25\textwidth}
         \centering
         \includegraphics[width=\textwidth]{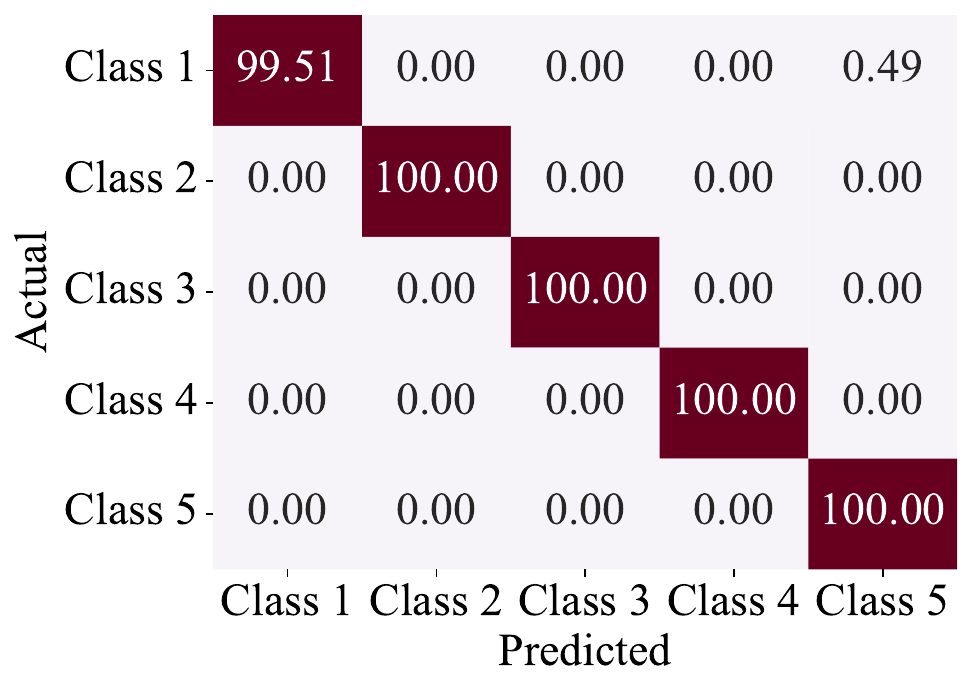}
         \caption{Client \#2}
         \label{fig:Heatmap2}
     \end{subfigure}
     \hspace{-8mm}
     \hfill
     \begin{subfigure}[t]{0.25\textwidth}
         \centering
         \includegraphics[width=\textwidth]{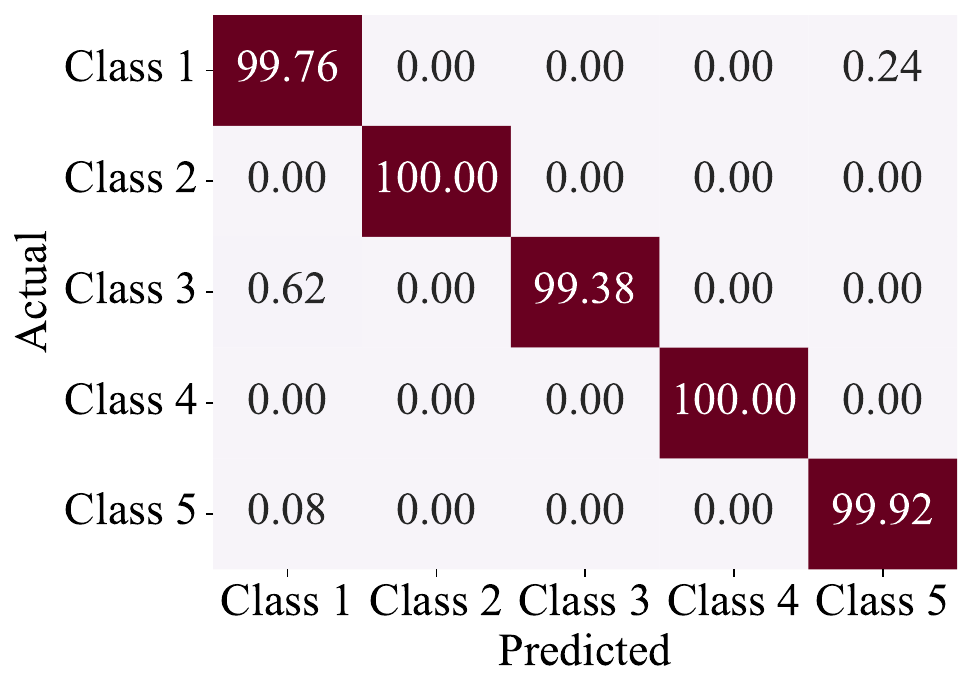}
         \caption{Client \#3}
         \label{fig:Heatmap3}
     \end{subfigure}
     \hspace{-8mm}
     \hfill
     \begin{subfigure}[t]{0.25\textwidth}
         \centering
         \includegraphics[width=\textwidth]{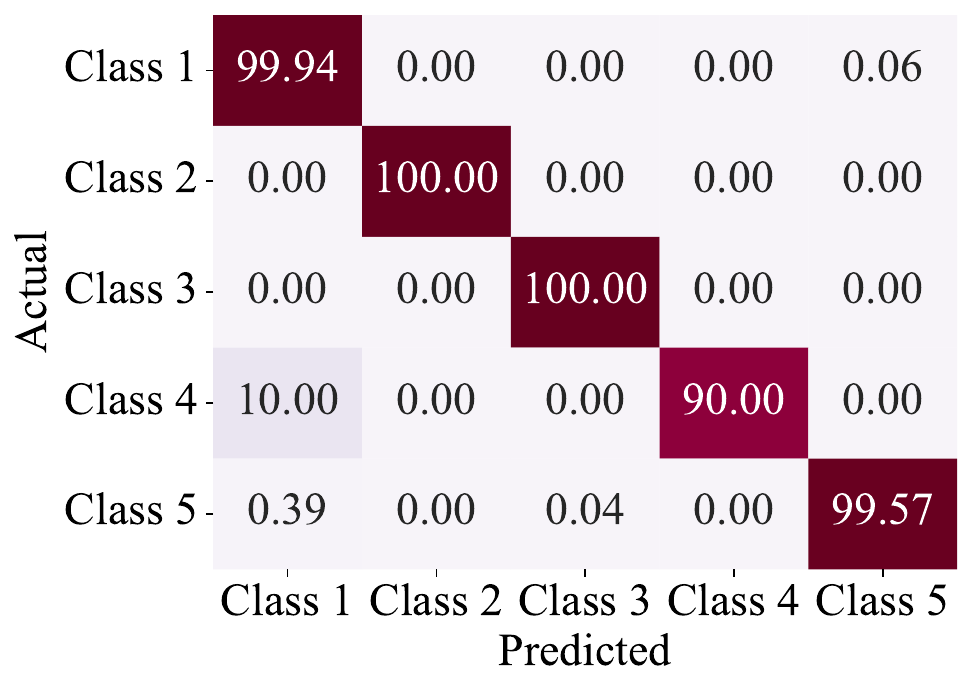}
         \caption{Client \#4}
         \label{fig:Heatmap4}
     \end{subfigure}
        \caption{Confusion matrix for each electric utility client.}
        \label{fig:Heatmaps}
    \hspace{-8mm}
\end{figure*}

All previous local TSA models based on the network architecture presented in subsection \ref{sec:architecture} are tested in this section. Table \ref{Tbl:ModelParams} summarizes the local and global training parameters of the TSA model. Figures \ref{fig:Heatmap1}, \ref{fig:Heatmap2}, \ref{fig:Heatmap3}, and \ref{fig:Heatmap4} show the prediction of the 4 local TSA models based on the proposed framework. One can see that most local client TSA models have predicted most classes perfectly. Some Class 4 data points were predicted as Class 1 for the 4th client and this has to do with the fact that Class 4 is an inflection class to both Class 1 or Class 5 (stable or unstable classes). In general, the proposed approach works effectively on the problem we are trying to solve. One can also observe that Class 1 and Class 5 are not 100\% perfect for most clients and that is because these two classes interchange quite a lot. 

\begin{figure*}
     \centering
     \begin{subfigure}[t]{0.25\textwidth}
         \centering
         \includegraphics[width=\textwidth]{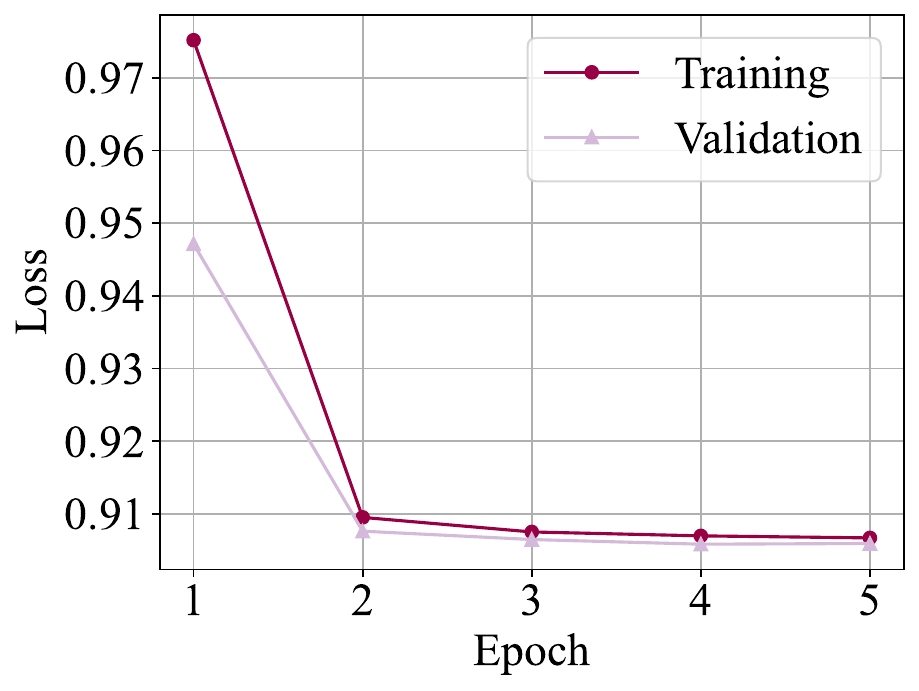}
         \caption{Client \#1}
         \label{fig:Loss1}
     \end{subfigure}
     \hspace{-8mm}
     \hfill
     \begin{subfigure}[t]{0.25\textwidth}
         \centering
         \includegraphics[width=\textwidth]{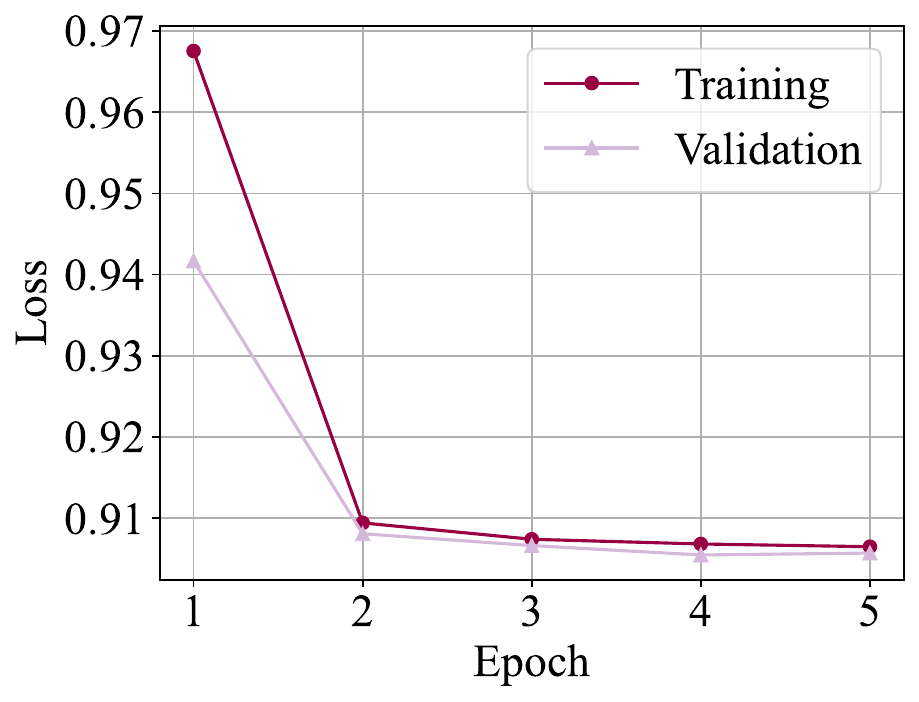}
         \caption{Client \#2}
         \label{fig:Loss2}
     \end{subfigure}
     \hspace{-8mm}
     \hfill
     \begin{subfigure}[t]{0.25\textwidth}
         \centering
         \includegraphics[width=\textwidth]{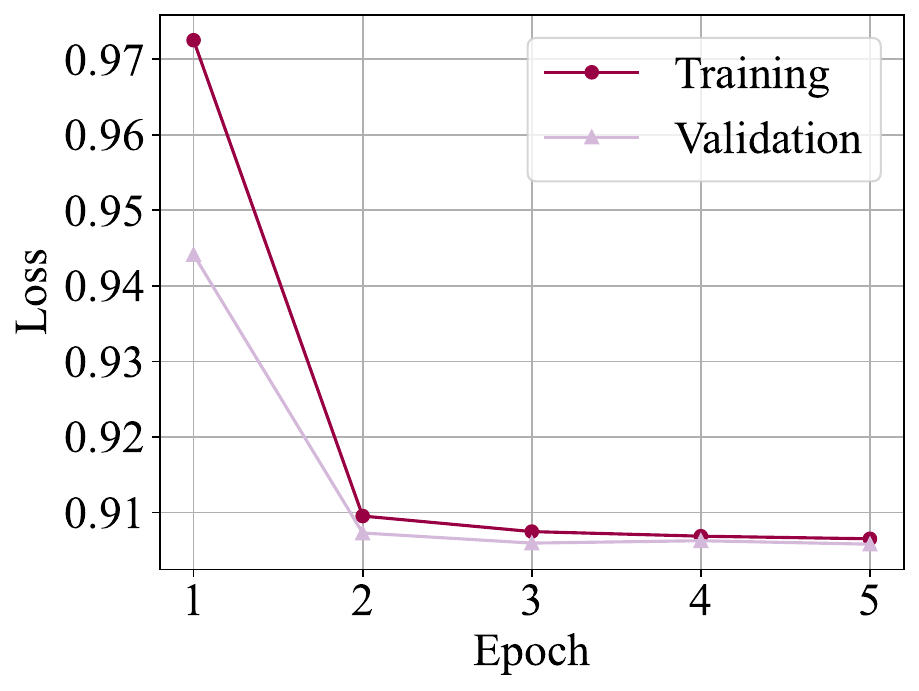}
         \caption{Client \#3}
         \label{fig:Loss3}
     \end{subfigure}
     \hspace{-8mm}
     \hfill
     \begin{subfigure}[t]{0.25\textwidth}
         \centering
         \includegraphics[width=\textwidth]{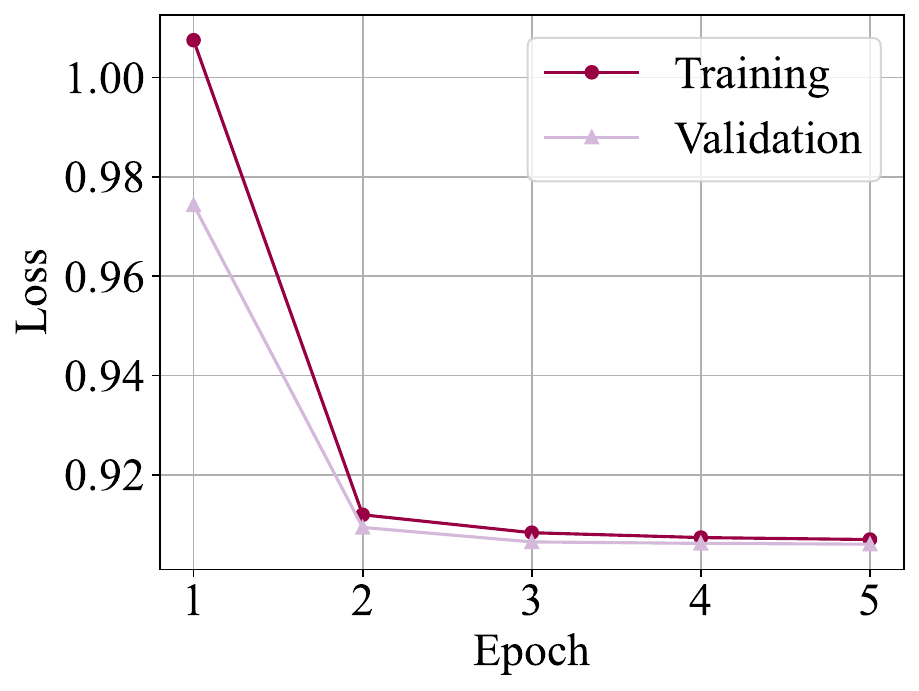}
         \caption{Client \#4}
         \label{fig:Loss4}
     \end{subfigure}
        \caption{Average training and validation losses for each communication round for all local epochs for each electric utility client.}
        \label{fig:Losses}
    \hspace{-8mm}
\end{figure*}

The training and validation losses for each communication round for all local training epochs for each utility client are shown in Figs. \ref{fig:Loss1}, \ref{fig:Loss2}, \ref{fig:Loss3}, and \ref{fig:Loss4}, and reveal that each local utility client model is not over-fitting and that the losses for each local utility client are decreasing.




\vspace{-4mm}
\section{Conclusion} \label{CON}
Centralized DL-based TSA models place a substantial reliance on computational power for their model training. Such centralized training schemes also pose a threat to the privacy of local utility data, as the data is transmitted to a central server for training. Consequently, these centralized models become susceptible to potential cyber-attacks and communication failures/delays. In response, this paper introduces a federated DL-based TSA model, where each local utility independently trains its own TSA model using its local dataset. This approach safeguards the privacy of data belonging to local utilities and demands less computational power compared to the centralized training scheme. The results obtained from testing this approach on a system composed of 4 identical IEEE 39-bus test systems, representing four distinct utilities with varying load conditions, as well as different fault scenarios, underscore the effectiveness and accuracy of this framework. It is also particularly adept at detecting intricate system operating states.
\vspace{-8mm}
\bibliography{reference} 
\bibliographystyle{ieeetr}
\end{document}